%% file: pp.tex
\documentclass[acmsmall, nonacm]{acmart}
\usepackage[utf8]{inputenc}
\usepackage{fontenc}
\pdfoutput=1
\makeatletter                   %1
\def\mdseries@tt{m}             %1
\makeatother                    %1
\usepackage[draft=true]{minted} %2
\usepackage{color}
\hypersetup{
    colorlinks=true,
    linkcolor=blue,
    filecolor=red,
    urlcolor=magenta,
    breaklinks=true,            %3
}
\usepackage{breakurl}           %3
\startPage{1}

\setcopyright{none}
\usepackage{multirow}

\usepackage{booktabs}   %% For formal tables:
\include{config}

\begin{document}
\sloppy                         %4
%% Title information
\title{Parallel Binary Code Analysis}         %% [Short Title] is optional;
                                        %% when present, will be used in
                                        %% header instead of Full Title.
%\titlenote{with title note}             %% \titlenote is optional;
                                        %% can be repeated if necessary;
                                        %% contents suppressed with 'anonymous'
%\subtitle{Subtitle}                     %% \subtitle is optional
%\subtitlenote{with subtitle note}       %% \subtitlenote is optional;
                                        %% can be repeated if necessary;
                                        %% contents suppressed with 'anonymous'

%% Author information
%% Contents and number of authors suppressed with 'anonymous'.
%% Each author should be introduced by \author, followed by
%% \authornote (optional), \orcid (optional), \affiliation, and
%% \email.
%% An author may have multiple affiliations and/or emails; repeat the
%% appropriate command.
%% Many elements are not rendered, but should be provided for metadata
%% extraction tools.

%% Author with single affiliation.
\author{Xiaozhu Meng}
                                        %% can be repeated if necessary
\orcid{nnnn-nnnn-nnnn-nnnn}             %% \orcid is optional
\affiliation{
  \department{Department of Computer Science}              %% \department is recommended
  \institution{Rice University}            %% \institution is required
  \city{Houston}
  \state{TX}
  \postcode{77005}
  \country{USA}                    %% \country is recommended
}
\email{Xiaozhu.Meng@rice.edu}          %% \email is recommended

\author{Jonathon M. Anderson}
                                        %% can be repeated if necessary
\orcid{nnnn-nnnn-nnnn-nnnn}             %% \orcid is optional
\affiliation{
  \department{Department of Computer Science}              %% \department is recommended
  \institution{Rice University}            %% \institution is required
  \city{Houston}
  \state{TX}
  \postcode{77005}
  \country{USA}                    %% \country is recommended
}
\email{jma14@rice.edu}          %% \email is recommended

\author{John Mellor-Crummey}
                                        %% can be repeated if necessary
\orcid{nnnn-nnnn-nnnn-nnnn}             %% \orcid is optional
\affiliation{
  \department{Department of Computer Science}              %% \department is recommended
  \institution{Rice University}            %% \institution is required
  \city{Houston}
  \state{TX}
  \postcode{77005}
  \country{USA}                    %% \country is recommended
}
\email{johnmc@rice.edu}          %% \email is recommended

\author{Mark W. Krentel}
                                        %% can be repeated if necessary
\orcid{nnnn-nnnn-nnnn-nnnn}             %% \orcid is optional
\affiliation{
  \department{Department of Computer Science}              %% \department is recommended
  \institution{Rice University}            %% \institution is required
  \city{Houston}
  \state{TX}
  \postcode{77005}
  \country{USA}                    %% \country is recommended
}
\email{krentel@rice.edu}          %% \email is recommended

\author{Barton P. Miller}
                                        %% can be repeated if necessary
\orcid{nnnn-nnnn-nnnn-nnnn}             %% \orcid is optional
\affiliation{
  \department{Computer Sciences Department}              %% \department is recommended
  \institution{University of Wisconsin-Madison}            %% \institution is required
  \city{Madison}
  \state{WI}
  \postcode{53706}
  \country{USA}                    %% \country is recommended
}
\email{bart@cs.wisc.edu}          %% \email is recommended

\author{Srđan Milaković}
                                        %% can be repeated if necessary
\orcid{nnnn-nnnn-nnnn-nnnn}             %% \orcid is optional
\affiliation{
  \department{Department of Computer Science}              %% \department is recommended
  \institution{Rice University}            %% \institution is required
  \city{Houston}
  \state{TX}
  \postcode{77005}
  \country{USA}                    %% \country is recommended
}
\email{sm108@rice.edu}          %% \email is recommended
\renewcommand{\shortauthors}{}
%% Abstract
%% Note: \begin{abstract}...\end{abstract} environment must come
%% before \maketitle command
\begin{abstract}
Binary code analysis is widely used to assess a program's correctness, performance, and provenance.
Binary analysis applications often construct control flow graphs, analyze data flow, and use debugging information to
understand how machine code relates to source lines, inlined functions, and data types.
To date, binary analysis has been single-threaded,
which is too slow for applications such as performance analysis and software forensics,
where it is becoming common to analyze binaries that are gigabytes in
size and in large batches that contain thousands of binaries.

%%%% johnmc: if there are thousands of binaries to analyze, analyze many
%%%% individual binaries in parallel using a single thread each.
%  .

This paper describes our design and implementation for accelerating the task of constructing control flow graphs (CFGs) from binaries with multithreading.
Existing research focuses on addressing challenging code constructs encountered during constructing CFGs, including
functions sharing code, jump table analysis, non-returning functions, and tail calls.
However, existing analyses do not consider the complex interactions between concurrent analysis of shared code,
making it difficult to extend existing serial algorithms to be parallel.
A systematic methodology to guide the design of parallel algorithms is essential.
We abstract the task of constructing CFGs as repeated applications of several core CFG operations regarding to
creating functions, basic blocks, and edges.
We then derive properties among CFG operations, including operation dependency, commutativity, monotonicity.
These operation properties guide our design of a new parallel analysis for constructing CFGs.
We achieved as much as 25$\times$ speedup for constructing CFGs on 64 hardware threads.
Binary analysis applications are significantly accelerated with the new parallel analysis:
we achieve 8$\times$ for a performance analysis tool and 7$\times$ for a software forensic tool with 16 hardware threads.
\end{abstract}

%% 2012 ACM Computing Classification System (CSS) concepts
%% Generate at 'http://dl.acm.org/ccs/ccs.cfm'.

%% End of generated code

%% Keywords
%% comma separated list
%\keywords{keyword1, keyword2, keyword3}  %% \keywords are mandatory in final camera-ready submission

%% \maketitle
%% Note: \maketitle command must come after title commands, author
%% commands, abstract environment, Computing Classification System
%% environment and commands, and keywords command.
\maketitle

\setminted{autogobble, fontsize=\scriptsize, breaklines,
  linenos, xleftmargin=1em}
\newmintinline[inC]{cpp}{fontsize=\normalsize}

\section{Introduction}

Binary code analysis is a foundational technique for a variety of applications,
including performance analysis~\cite{HPCToolkit2010,Tapus2002AcitveHarmony,Paradyn1995}, software correctness~\cite{Arnold2007STAT, Gu2018DDR},
software security~\cite{Jacobson14ReuseAttack,vanderVeen2015PCC,vanderVeen2016ATC}, and software forensics~\cite{Rosenblum2011BinAuthor,Meng2017MultiAuthor}.
Important binary code analysis capabilities include
constructing control flow graphs (CFGs), analyzing control flow and data flow properties,
and extracting source line mappings and data types from debugging information, when it is available.
Traditionally, binary analysis applications are single-threaded.
However, recent trends in these applications call for improving the performance of binary analysis applications.

In the field of performance analysis,
it is becoming common to optimize the performance of large software systems that compile into multi-gigabyte binaries.
We have witnessed this trend within software developed by national
laboratories and popular
machine learning frameworks such as TensorFlow~\cite{tensorflow}.
The developers of these large softwares use
the following performance analysis workflow to optimize their code:
(1) compile the source code to generate the binary program,
(2) measure the performance of the binary during execution,
(3) attribute measurements to the corresponding source code (via binary analysis),
and (4) optimize the source based on the performance results.
These four steps are repeated until developers are satisfied with
their software's performance.

In this performance analysis cycle, binary analysis must be repeated after any source code change
because even small code changes can lead to dramatically different binaries, especially with
C++ templates and aggressive compiler optimizations.
In such a workflow, if the binary analysis in step (3) is slow,
it will reduce the throughput and effectiveness of performance analysis.
Current single-threaded binary code analysis takes too long to analyze such large binaries.
It takes more than 20 minutes to analyze a 7.7GiB shared library from TensorFlow,
which would interrupt the workflow of developers tuning the code for production.

In the field of software forensics, researchers have achieved great success in applying machine learning to tasks
including compiler identification~\cite{Rosenblum2011Toolchain} and
authorship attribution~\cite{caliskan2018coding,Meng2017MultiAuthor,Rosenblum2011BinAuthor}.
These machine learning-based software forensics applications require large training sets to be effective,
containing hundreds to thousands of binaries.
During the development of these forensic applications, the developers typically repeat the following workflow:
(1) design a set of binary code features,
(2) extract the features with binary analysis to construct a training set,
(3) validate the accuracy of a model trained using the new features.
These three steps are repeated until the developers are satisfied with the effectiveness of the binary code features.

While the training and tuning of machine learning models have traditionally been
regarded as the bottleneck of these software forensics applications,
modern machine learning packages often provide support for parallel training and tuning,
using multithreading and even GPUs.
In such scenario, a serial feature extraction step can be a limiting factor of the development cycle:
for example, the feature extraction step in compiler identification~\cite{Rosenblum2011Toolchain} may take over 24 hours.

In this paper, we present our design and implementation of parallel binary code analysis to address
the speed requirements imposed by these applications.
The core of this work is a new parallel analysis for constructing control flow graphs (CFG construction),
which constructs functions, basic blocks, and edges between basic blocks.
CFG construction is used in nearly every binary analysis application, directly or indirectly.

Modern serial CFG construction algorithms focus on understanding complex machine code generated by compilers~\cite{Meng2016BinaryNotEasy, DiFederico2017RUB, Angr2016}.
Complex code constructs such as non-returning functions, tail calls and jump tables play key
roles in understanding high-level programming constructs, making their analysis important for
applications. While function level parallelism is a natural starting point for parallel CFG construction,
we must address a range of complex issues:

\begin{itemize}
\item Functions may share code. Threads analyzing different functions may end up concurrently analyzing shared code and require synchronization.
\item Control flow graphs evolve during analysis. As a result, a parallel algorithm for CFG construction needs to consider concurrent changes by others.
\item Current binary code analysis is not designed nor implemented with parallelism in mind. Parallelization exposes the flaws in existing serial analysis for jump tables and tail call identification.
\item While protecting intricate data structures with mutual exclusion is a tempting way to guarantee correctness, the serialization this induces must be carefully evaluated for its impact on parallelism and performance.
\end{itemize}

To systematically address these issues, we abstract CFG construction as repeated applications of several primitive CFG construction operations.
These operations include creation of CFG elements such as functions, basic blocks, edges, modification to basic block ranges, and removing blocks and edges.
We derive operation properties, including dependencies, commutativity, and monotonicity, and
use this theoretical framework to reason the correctness and performance of CFG construction algorithms.
This abstraction allows us to identify flaws in existing serial CFG construction algorithms.
Many of the flaws are caused by not considering the interactions between complex code constructs.
This methodology enables us to express parallelism as commutative operations and focus our attention to address operation dependencies to improve performance.
We then design new algorithms and data structures for parallel CFG construction to address both correctness and performance issues.

%(\cref{sec:methodology}).
%The final design of our parallel binary code analysis (\cref{sec:parseapi}) and parallel DWARF parsing (\cref{sec:symtabapi}),
%and the implementation of our new analyses (\cref{sec:implementation}) are the results of iteratively applying this methodology.

We implemented our new parallel CFG construction in the Dyninst binary analysis and instrumentation toolkit~\cite{DyninstAPI},
a library widely used by researchers in performance analysis, security, and software forensics, and
evaluated the performance characteristics of our parallel binary analysis with a number of large binaries,
including a 7.7GiB shared library from TensorFlow.
We achieved as much as 25$\times$ speedup for constructing control flow graphs on 64 hardware threads, which significantly accelerates client tools that employ binary analysis.
We then showcase the benefits of parallel binary analysis with two applications.
The \texttt{hpcstruct} utility in HPCToolkit~\cite{HPCToolkit2010} is used to relate performance measurement back to source code;
we achieved 8$\times$ speedup for \texttt{hpcstruct}.
\textit{BinFeat}~\cite{BinFeat} is a tool for extracting binary code features for software forensics, for which we achieved 7$\times$ speedup.

In summary, this work makes the following contributions:

\begin{enumerate}
\item A set of CFG operations and operation properties that enable us to reason correctness and performance of CFG construction algorithms.
\item A new algorithm for parallel CFG construction that is derived from the requirements and properties of CFG operations.
\item An implementation of the new algorithm in Dyninst that can be used by other binary analysis application developers.
\item Demonstrating the effectiveness of our new parallel binary analysis with two binary analysis applications: \texttt{hpcstruct} which significantly accelerates program structure recovery for performance analysis and \texttt{BinFeat} which significantly speeds up binary code feature extraction for software forensics.
\end{enumerate}

\section{Related Work}

There is rich literature about constructing CFGs from binaries~\cite{Kinder2012ACF, Bardin2011RCR, Kinder2008Jakstab, Schwarz2002Disassembly,DiFederico2017RUB}.
A commonly used approach is control flow traversal~\cite{Schwarz2002Disassembly, Theiling2000ESP}.
Starting from known function entry points such as the ones found in the symbol table,
it follows control flow transfers in the program to discover code and identify additional function entry points for further analysis.
We discuss several challenging code constructs that must be addressed during control flow traversal
and representative binary analysis tools that implement control flow traversal.

\subsection{Challenging Code Constructs}

\textbf{Functions sharing code:}
A common compiler optimization is to share binary code between functions with common
functionality, such as error handling code and stack tear-down code.
This construct is common in compiled code. We have observed this construct in glibc-2.29 (Released Jan. 2019)
where common error handling code is shared by multiple syscall wrappers,
and within code compiled by the Intel Compiler Suite (ICC).
This can also occur logically in functions with multiple entry points: binary analysis tools typically
represent such functions as multiple single-entry functions that share code.
Thus, Fortran functions with multiple programmer-specified entry points (using the \texttt{entry}
keyword), and binaries on Power 8 or newer (the ABI specifies that each function has at least
two entry points) lead to functions sharing code.
To support code shared multiple functions,
one can define a function as the basic blocks that are reachable from the function entry by traversing only intra-procedural edges~\cite{Bernat2012CFGEditing, Meng2016BinaryNotEasy}.

\textbf{Non-returning functions:}  Binary analysis tools often define a call fall-through edge, which is a summary edge representing that the control flow at a function call will return to the call site.
However, a function call to a non-returning function will never return to its call site,
so there should be no call fall-through edge at such call sites.
A wrongly created call fall-through edge can lead to confusing control flow and cascading impacts on binary analysis applications.

The general idea of identifying non-returning functions is to match function names against known non-returning functions such as \inC{exit} and \inC{abort} and
uses an iterative analysis to identify functions that always end in calls to non-returning functions.
One example is a fixed point analysis presented by Meng and Miller~\cite{Meng2016BinaryNotEasy}.
Each function has a return status, with three different values: UNSET, RETURN, and NORETURN.
A function's return status is initialized with NORETURN if it is known to be a non-returning function,
otherwise a function's return status is UNSET.
Three main components in the non-returning function analysis are:
(1) a function's return status is set to RETURN if we find a return instruction;
(2) if we encounter a call site calling to a function with UNSET return status,
we do not parse the call fall-through edge until the callee's return status is set to RETURN;
(3) if there is a cyclic dependency between functions' return statuses, all functions in the cycle are non-returning.

\textbf{Jump table analysis:} Compilers often emit indirect jumps for switch statements in the source code.
The targets of these indirect jumps are calculated based on jump table data in the binary.
Being able to statically resolve the control flow targets calculated through jump tables is critical for complete control flow traversal.
A common approach for resolving jump table targets is to use backward slicing to identify the instructions that
are involved in the target calculation and construct a symbolic expression of the jump target to identify the actual jump targets~\cite{Meng2016BinaryNotEasy, Angr2016, DiFederico2017RUB, WilliamsKing2020Egalito}.

\textbf{Tail calls:} A tail call~\cite{Clinger98propertail} is a compiler optimization that uses a jump instruction at the
end of a function to target the entry point of another function, thus not every branch
should be labeled as intra-procedural. Tail calls are often recognized through heuristics~\cite{Meng2016BinaryNotEasy, DiFederico2017RUB}, including
(1) a branch to a known function entry is a tail call;
(2) a branch to a basic block that is reachable through only intra-procedural edges of the current function is not a tail call;
(3) if there is stack frame tear down before the branch, it is a tail call.

\subsection{Binary Analysis Tools}

Recent binary analysis tools address these challenging code constructs,
including angr~\cite{Angr2016}, Dyninst~\cite{Meng2016BinaryNotEasy}, and rev.ng~\cite{DiFederico2017RUB}.
While these tools share similarity in addressing challenging code constructs,
the software infrastructure of these tools have distinct characteristics regarding to analysis speed.

Both angr and rev.ng first lift machine instructions to IR and then perform analysis on the resulting IR.
The first advantage of this approach is that the binary analysis is not architecture specific and can be readily ported to a new architecture
after the IR is supported on the new architecture.
For example, rev.ng uses QEMU to lift binary to LLVM IR. Therefore, rev.ng supports every architecture where QEMU is supported (more than 16 different architectures).
Similarly, angr uses Valgrind's VEX IR.
The second advantage is that lifting to IR facilitates the development of complex data flow analysis such as points-to analysis and value set analysis.
However, this approach leads to significant performance slowdown for two reasons.
First, lifting process itself is slow.
Second, The number of assignments in the IR is significantly larger than the number of machine instructions as one instruction may be lifted to multiple IR assignments,
especially on CISC architectures such as x86-64.

On the other hand, Dyninst directly operates with the binary.
Dyninst's instructionAPI provides an architecture independent interface for querying instruction opcodes,
instruction operands, registers, and memory addressing modes.
The CFG construction code inside Dyninst works with this ``bare-metal'' instruction interface.
The only exception is that when Dyninst resolves jump tables, Dyninst lifts machine instructions to ROSE IR~\cite{quinlan2011rose}.
However, since lifting is applied to instructions that are involved in the jump table calculation found by backward slicing,
typically only a small portion of the binary is lifted.

As we will describe in~\cref{sec:application}, complex data flow analysis are not needed in our target applications.
Therefore, we implement our new parallel CFG construction algorithms in Dyninst to achieve better performance.

\section{Notation} \label{sec:notation}

\newcommand\tuple[1]{\langle #1 \rangle}
\newcommand\edge[2]{(#1 \rightarrow #2)}

While existing literature comprehensively describes how to address each challenging code construct,
a critical problem we encountered when designing a parallel CFG construction algorithm is the complex interactions between these code constructs.
Serial algorithms are designed with the assumption that no concurrent modification is made to the CFG.
To address this problem, we present an abstraction of control flow graphs and a series of core operations on them.
This abstraction  enables us to reason interactions between different CFG construction operations.

Our abstraction builds upon the CFG definitions and operations designed for binary modification~\cite{Bernat2012CFGEditing}, which mainly works with fully constructed CFGs.
Our abstraction instead focuses on the process of constructing CFGs.

\textbf{Definitions:} We define a CFG $G = \tuple{B,C,E,F}$ to be a tuple of the following:
\begin{itemize}
\item $B$ is a set of address ranges $[s,e)$, representing basic blocks within the binary. Each of these contains at most one control flow instruction, which if present is the final instruction within the range, and has incoming control flow at only address $s$.
\item $C$ is a set of candidate blocks $[t]$, representing addresses which are known to start basic blocks but do not have known ending addresses yet.
\item $E\subseteq \{\edge{a}{b} : a\in B, b\in B\cup C \}$ is a set of directed edges between basic blocks, representing possible control flow executions between blocks.
\item $F\subseteq B\cup C$ is the set of function entry blocks.
\end{itemize}

\textbf{Partial order}: We utilize a partial order between control flow graphs, designed such that a larger graph includes more control flow elements. We define $G_1\preccurlyeq G_2$ if all of the following are true:
\begin{itemize}
\item The address ranges contained in $G_1$ are also contained by $G_2$. Formally, let $A_1$ and $A_2$ be the addresses contained by the blocks in $B_1$ and $B_2$ respectively. Then we require $A_1\subseteq A_2$.
\item The explicit control flow present in $G_1$ is also present in $G_2$, regardless of adjustments to block ranges. Formally, for every edge $\edge{a=[s_a,e_a)}{b=[s_b,e_b)}$ or $\edge{a=[s_a,e_a)}{b=[s_b]}\in E_1$, one of the similar edges $\edge{[s'_a,e_a)}{[s_b,e'_b)}$ and $\edge{[s'_a,e_a]}{[s_b]}$ must be present in $E_2$. Intuitively, $G_2$ may contain additional control flow edges that target addresses inside $a$ or $b$, causing them to be split. The requirement here is that the end address of the source block $e_a$ and the start address of the target block $s_b$ are preserved under the partial order.
\item The implicit control flow through a basic block in $G_1$ is preserved in $G_2$. Formally, for every block $b=[s_0,e)\in B_1$ there is a sequence of blocks $[s_0,s_1),\dots,[s_n,e)\in B_2$ such that $\edge{[s_i,s_{i+1})}{[s_{i+1},s_{i+2})}\in E_2$ for $i=0,\dots,n-2$. This means that a block $b$ in $G_1$ can be split into multiple smaller blocks in $G_2$ to incorporate other incoming control flow.
\item Function entry labels in $G_1$ are preserved in $G_2$, regardless of range adjustments. Formally, for every block $[s,e)$ or $[s]\in F_1$, there is a block starting at the same address $[s,e')$ or $[s]\in F_2$.
\end{itemize}

\textbf{CFG operations}: To construct a CFG based on the underlying binary, we define a number of core operations:
\begin{itemize}

\item Block End Resolution: Given a graph $G$ containing a candidate block $[t]\in C$, we define $O_{BER}(G, [t])$ as $G$ with the candidate block $[t]$ replaced by an actual basic block starting at $t$ with a determined end address. There are three possible cases:
\begin{itemize}
\item Block splitting. If there is an existing block $b = [s,e)\in B$ such that $s < t < e$, then we have to split $b$ into the basic blocks $[s,t)$ and $[t,e)$. Any incoming edges on $b$ are redirected to $[s,t)$, while outgoing edges on $b$ and incoming edges on $[t]$ are moved to $[t,e)$.
\item Early block ending. If there is an existing block $b = [s,e)\in B$ such that $t < s$ and the range $[t,s)$ contains no control flow instructions, we replace $[t]$ with $[t,s)$ as in the first case and append the edge $\edge{[t,s)}{[s,e)}$.
\item Linear parsing. If neither of the previous cases apply, let $e$ be the address directly after the first control flow instruction following $t$. We replace $[t]$ with $[t,e)$ as in the first case.
\end{itemize}

\item Direct Edge Creation: Given a block $a$ in a graph $G$, which ends with a direct control flow instruction, we define $O_{DEC}(G,a)$ as $G$ with
outgoing edges appended to $a$, based on the control flow instruction within $a$ (if one exists). There are three cases:
\begin{itemize}
\item If $a$ terminates with an unconditional jump to address $t$, we append the edge $\edge{a}{[t]}$.
\item If $a = [s,e)$ terminates with a conditional jump to address $t$, we append edges for the cases where the condition is true $\edge{a}{[t]}$ and false $\edge{a}{[e]}$.
\item If $a$ terminates with a function call instruction to address $t$, we append the edge $\edge{a}{[t]}$.
\end{itemize}

\item Call Fall-Through Edge Creation: Given an edge $e = \edge{[s,e)}{f}$ in a graph $G$ where $[s,e)$ contains a function call instruction and $f\in F$, we define $O_{CFEC}(G,e)$ as $G$ potentially with the additional edge $\edge{[s,e)}{[e]}$ summarizing the execution of the callee function. Correct application of this operation depends on the non-returning function analysis used to identify whether the target function can return or not.

\item Indirect Edge Creation: Given a block $a$ in a graph $G$ which contains a jump to a dynamic address, we define $O_{IEC}(G,a)$ as $G$ with the additional edges $\edge{a}{[t_1]},\dots,\edge{a}{[t_n]}$, where $t_1,\dots,t_n$ are target addresses determined statically. It is possible for this operation to add no edges if the analysis used is insufficient to statically determine the possible targets.

\item Function Entry Identification: Given an edge $e = \edge{a}{b}$ in a graph $G$, we define $O_{FEI}(G, e)$ as $G$ with the block $b$ potentially labeled as a function entry.
This operation is trivial if $e$ was created by an explicit call instruction, but further heuristics are required to identify functions that are reached only through optimized tail calls.

\item Edge Removal: Given an edge $e = \edge{a}{b}$ within a graph $G$, we define $O_{ER}(G,e)$ as $G$ with the edge $e$ removed along with any blocks and edges that are no longer reachable from any function entry point. Formally, let $B'\subseteq B$ and $C'\subseteq C$ be the sets of blocks and candidate blocks in $G$ reachable from any block in $F$ without traversing $e$. We can then define
\begin{equation*}
  O_{ER}(G,e) = \tuple{B', C', E\cap\{\edge{a'}{b'} : a'\in B', b'\in B'\cup C'\}\setminus \{e\}, F}.
\end{equation*}

\end{itemize}

Starting with the initial graph $G_0=\tuple{\varnothing, F_0, \varnothing, F_0}$, where $F_0$ is the set of candidate function entry blocks discovered via the binary's symbol table and unwind
information, the task of CFG construction can be abstracted as repeated application of these operations. We denote $G_1, G_2 \cdots, G_{n-1}$ as the intermediate results and $G_n$ as the final CFG.

\section{CFG Operation Properties} \label{sec:properties}

We present several important properties of the defined operations, assess existing serial algorithms with these properties, and use these properties to define critical correctness and performance issues for parallel CFG constructions.

\subsection{Properties}

\textbf{Operation dependencies:} To correctly build the CFG, operations should be applied with an order that satisfy the dependencies among them. We identify two types of dependencies:
\begin{itemize}
  \item \textbf{Applicability Dependency}. We cannot apply operations to an graph element that has not been discovered. For example, we must create an edge before we can resolve the target block candidate of this edge.
  \item \textbf{Non-returning Function Dependency}. The correctness of $O_{CFEC}$ for creating call fall-through edges depends on the operations applied to the callee functions to determine whether the callee would return or not. If $O_{CFEC}$ is applied when the callee does not return, an erroneous call fall-through edge would be added, leading to an incorrect CFG.
\end{itemize}

Operations that satisfy either type of the above dependency must be applied in order. We classify operations that are not constrained by any dependency into three categories:

\begin{itemize}
\item \textbf{Commutative operations}: The operations $O_{BER}$ and $O_{DEC}$ commute with themselves and with each other, allowing us to choose an order convenient for processing. To establish this, we discuss the following three cases:
\begin{itemize}
\item Given two candidate blocks $[a]$ and $[b]$ where $a < b$, we have $O_{BER}(O_{BER}(G, [a]), [b]) = O_{BER}(O_{BER}(G, [b]), [a])$. First, if there is a control flow instruction ending at address $c$ where $a < c < b$, candidate block $[a]$ will end before $c$ while candidate block $[b]$ will end after $c$. These two operations will act on non-overlapping address ranges and be independent, which gives us commutativity. Second, if a control flow instruction ends at $c$ where $a < b < c$ and $c$ is first control flow instruction following $b$, we have

$
\begin{aligned}
  O_{BER}(O_{BER}(G, [a]), [b]) & = O_{BER}( G \cup \{[a, c)\} , [b])  & \text{(Linear parsing)} \\
                                & = G \cup \{ [a, b), [b, c) \}   & \text{(Block splitting)}  \\
                                & = O_{BER}(G \cup \{ [b, c) \}, [a])  & \text{(Early block ending)} \\
                                & = O_{BER}(O_{BER}(G, [b]), [a]). & \text{(Linear parsing)}
\end{aligned}
$

Thus we also have commutativity in this case.

\item Given two blocks $a$ and $b$, we have $O_{DEC}(O_{DEC}(G, a), b) = O_{DEC}(O_{DEC}(G, b), a)$. This is because $O_{DEC}(G, a)$ only considers the terminating control flow instructions within the block $a$.
\item We have $O_{BER}(O_{DEC}(G,[s,e)),[t]) = O_{DEC}(O_{BER}(G,[t]),[s,e))$, when given a candidate block $[t]$ and a block $[s,e)$. We observe that $O_{DEC}(G, [s,e))$ depends on only the terminating control flow instruction ending at $e$ and will generate only new candidate blocks while $O_{BER}(G, [t])$ does not depend on candidate blocks. Therefore these two operations are independent and thus commutative.
\end{itemize}

The operation $O_{ER}$ also commutes with itself, allowing us to choose an order convenient for processing. The graph $O_{ER}(O_{ER}(G, e_1), e_2) = O_{ER}(O_{ER}(G, e_2), e_1)$ will contain no blocks reachable only through $e_1$ and $e_2$, which gives us the commutativity property.

\item \textbf{Monotonic ordering property}: While $O_{IEC}$ does not commute trivially with any other operation, we can still establish a weaker property. Let $O_{IEC}(G, a)$ be an indirect edge creation operation and $O_x$ be an $O_{BER}$ or $O_{DEC}$ operation. We have $O_x(O_{IEC}(G, a)) \preccurlyeq O_{IEC}(O_x(G), a)$, since the edges added by $O_x$ can at most add control flow paths preceeding $a$ and thus increase the set of target addresses. When our goal is to achieve a maximal CFG, this allows us to reorder $O_{IEC}$ after any $O_{BER}$ and $O_{DEC}$ operations without decreasing the final result.

\item \textbf{Non-reorderable operations}: The operations $O_{CFEC}$ and $O_{FEI}$ do not always commute, nor do they satisfy the ordering property above in all cases.
Both of these operations use implementation-specific analyses: non-returning function analysis for $O_{CFEC}$ and tail call identification heuristics for $O_{FEI}$, both of which at times require inspection of large portions of the graph. Because of the sensitivity of these operations, we are cautious to apply these operations only after the considered subgraph has stabilized.
\end{itemize}

\subsection{Serial Algorithm Assessment} \label{sec:property:serial}

We compare the serial algorithms used by angr~\cite{Angr2016}, Dyninst~\cite{Meng2016BinaryNotEasy}, and rev.ng~\cite{DiFederico2017RUB} using the defined notation and operations.

\begin{listing}[t]
  \begin{minted}[escapeinside=||, mathescape=true]{gas}
  A:            B:
  ...           ...
  leaveq        mov %rsi, 1
  jmp 0x400     jmp 0x400
  \end{minted}
  \caption{\small An example showing inconsistent results in the tail call heuristics used by Dyninst.}
  \label{fig:tailcall}
\end{listing}

\begin{itemize}
\item Dyninst and angr's CFG construction can be characterized with an increasing expression: $G_0 \preccurlyeq G_1 \preccurlyeq G_2 \cdots \preccurlyeq G_n$.
This increasing pattern has the advantage of not performing redundant work of adding and then removing graph elements.
On the other hand, rev.ng has an additional step to clean candidate function entries after this increasing phase.

We observe that this cleaning step can address issues caused by non-commutating operations such as tail call identification.
\cref{fig:tailcall} is an example where Dyninst will give inconsistent results depending on the analysis order.
In this example, function A and B branch to the same address.
If A is analyzed first, because \inC{leaveq} tears down the stack frame, Dyninst will treat the branch in A as a tail call and create a new function at the branch target;
later when Dyninst analyze B, we find that B branches to a known function entry, so the branch is in B also a tail call.
In this case, function B will not include block at 0x400.
On the other hand, if B is analyzed first, because there is no stack frame tear down before the branch in B, Dyninst will not treat the branch as a tail call, and the block at 0x400 will be part of B.
Therefore, the function boundary of B is determined by the order of analysis.
Note that without other context, it is equally valid to conclude either ``A and B both tail call to 0x400'' or ``A and B share block at 0x400''.
With a cleaning step at the end, we have the opportunity to generate a consistent answer.

\item The jump table analysis implemented in tools does not necessarily satisfy the monotonicity property we defined for $O_{IEC}$.
The root cause of this issue is imperfect jump table analysis where jump table targets can be over-approximated.
Suppose we have $O_{IEC}(G, b_1)$ and $O_{IEC}(G, b_2)$.
Due to imperfect jump table analysis, $O_{IEC}(G, b_1)$ generates an over-approximated set of jump targets, resulting in invalid outgoing edges.
Such additional but confusing control flow may cause $O_{IEC}(G, b_2)$ to fail, leading an empty set of targets.
However, if $O_{IEC}(G, b_2)$ is performed first, then we may get the correct non-empty set of jump targets for $b_2$.
We have observed this problem in Dyninst's jump table analysis.
While rev.gn and angr both provide detailed descriptions on how they resolve jump tables, neither is able to guarantee no over-approximation of the jump targets.
\end{itemize}

\subsection{Challenges Towards Parallelism}

Besides the need of a cleaning step after the increasing CFG construction and the need to address jump table over-approximation,
we further identify three issues must be addressed to achieve effective parallel analysis.

\begin{itemize}
\item Commutative operations still need careful synchronization.
Suppose we have two direct edges $e_1$ and $e_2$ that have the same target address and we perform $O_{DEC}(G, e_1)$ and $O_{DEC}(G, e_2)$ concurrently.
Due to commutativity, $O_{DEC}(O_{DEC}(G, e_1), e_2) = O_{DEC}(O_{DEC}(G, e_2), e_1)$.
However, the operation performed first will create the candidate block, making the second operation effectively the identity function.
This is trivial to maintain for serial algorithms, but careful synchronization is necessary to maintain this uniqueness property in a parallel setting.

\item Non-returning function dependencies between operations can also lead to ineffective parallelism. In a call chain where $F_1$ calls $F_2$, $F_2$ calls $F_3$, $\cdots$, and $F_{n-1}$ calls $F_n$,
an $O_{CFEC}$ operation in $F_1$ may need to wait for operations in $F_2$ to complete, which may need to wait for operations in $F_3$, and so on.
This effect causes undesirable serialization during the analysis.

\item The monotonic ordering property for operation $O_{IEC}(G,a)$ indicates that we might be able to find more control flow targets if it is applied after other edge creation operations.
However, deferring $O_{IEC}(G,a)$ can exacerbate the issue of non-returning function dependencies,
as this will delay the discovery of returns that are only reachable through the indirect jump.

\end{itemize}

\section{Parallel CFG Construction} \label{sec:parseapi}

In~\cref{sec:properties}, we have established that commutative operations can be performed in any order without impacting the final results.
This is the foundation for parallel CFG construction.
\Cref{sec:pp:stages} describes the stages of our parallel analysis.
\Cref{sec:pp:invariants} presents five invariants for supporting concurrent CFG operations.
\Cref{sec:pp:cft} discusses parallel control flow traversal.
Finally, \Cref{sec:pp:finalization} discusses the parallel finalization step that is needed to obtain a correct CFG.

\subsection{Stages in Parallel CFG Construction}
\label{sec:pp:stages}

Our parallel analysis can be characterized with the expression  $G_0 \preccurlyeq G_1 \preccurlyeq G_2 \cdots \preccurlyeq G_m \succcurlyeq G_{m+1} \succcurlyeq \cdots \succcurlyeq G_n$.
It contains a CFG expansion phase to discover and add new graph elements, followed by a CFG correction phase to remove incorrect graph components.

\Cref{fig:top_level_algo} describes three main stages in our parallel analysis.
It starts with initializing data structures for analyzing functions defined in the symbol table (Line 1).
It is necessary to parallelize this step as we have seen large binaries containing millions of functions.
The second stage represents the increasing construction phase.
We perform control flow traversal for initialized functions in parallel,
during which we may discover more functions.
We repeatedly apply control flow traversal until there are no more functions to analyze (Line 2 - 6).
The details of control flow traversal are presented in \cref{sec:pp:cft}.
The final stage is to finalize the CFG (Line 7).
This stage includes cleaning control flow edges and blocks created by over-approximated jump tables,
cleaning inconsistent tail call identification results, and determining which basic blocks belong to which function by
traversing intra-procedural edges from function entry blocks.

\begin{listing}[t]
\begin{minted}[escapeinside=||, mathescape=true]{moon}
  funcs = InitFunctions() -- Done in parallel
  while funcs != |$\emptyset$|
    more_funcs = |$\emptyset$|
    parallel for f in funcs
      more_funcs = more_funcs |$\cup$| ControlFlowTraversal(f)
    funcs = more_funcs
  CFGFinalization() -- Done in parallel
\end{minted}
\caption{\small Stages of our parallel binary analysis.}
\label{fig:top_level_algo}
\end{listing}

\subsection{Control Flow Traversal Invariants}
\label{sec:pp:invariants}

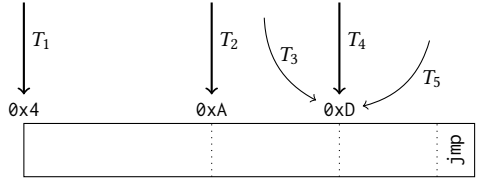
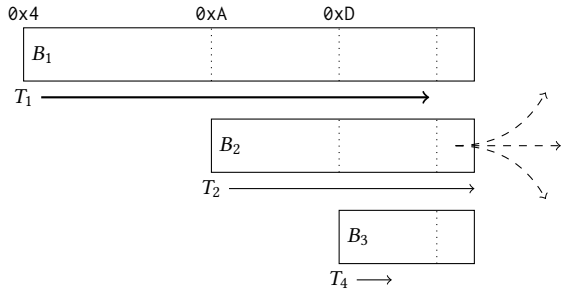
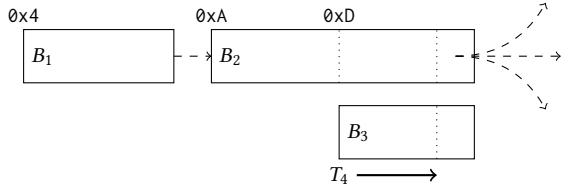
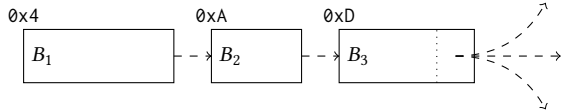
\begin{figure}
  \footnotesize
  \begin{subfigure}{.9\textwidth}
    \centering
    \begin{tikzpicture}
      \draw (0,0) rectangle (6,.7);
      \draw[dotted] (2.5,0) -- +(0,.7) (4.2,0) -- +(0,.7) (5.5,0) -- +(0,.7);
      \draw[above] (0,.7) node(X4) {\texttt{0x4}} +(2.5,0) node(Xa) {\texttt{0xA}}
        +(4.2,0) node(Xd) {\texttt{0xD}};
      \draw[<-,thick] (X4) -- +(0,1.4) coordinate(T2);           \draw (T2) +(0,-.5) node[right]{$T_1$};
      \draw[<-,thick] (Xa) -- +(0,1.4) coordinate(T4);           \draw (T4) +(0,-.5) node[right]{$T_2$};
      \draw[<-] (Xd) to[bend left,<-] +(-1,1.2) coordinate(T1);  \draw (T1) +(.1,-.5) node[right]{$T_3$};
      \draw[<-,thick] (Xd) -- +(0,1.4) coordinate(T5);           \draw (T5) +(0,-.5) node[right]{$T_4$};
      \draw[<-] (Xd) to[bend right,<-] +(1.2,.9) coordinate(T3); \draw (T3) +(-.2,-.5) node[right]{$T_5$};
      \path (5.75,.35) node[rotate=90]{\texttt{jmp}} -- +(1.4,0);
    \end{tikzpicture}
    \caption{$T_1$ and $T_2$ branch to two different addresses. $T_3$, $T_4$, and $T_5$ branch to the same target.} \label{fig:inv:sub1}
  \end{subfigure}
  \begin{subfigure}{.9\textwidth}
    \centering
    \begin{tikzpicture}
      \draw (0,2.4) node[below](T2){$T_1$} rectangle +(6,.7) +(0,.35) node[right]{$B_1$}
          +(0,.7) node[above]{\texttt{0x4}}
        (2.5,1.2) node[below](T4){$T_2$} rectangle +(3.5,.7) +(0,.35) node[right]{$B_2$}
        (4.2,0) node[below](T5){$T_4$} rectangle +(1.8,.7)   +(0,.35) node[right]{$B_3$};
      \draw[dotted] (2.5,2.4) -- +(0,.7) node[above]{\texttt{0xA}}
        (4.2,2.4) -- +(0,.7) node[above]{\texttt{0xD}} (5.5,2.4) -- +(0,.7)
        (4.2,1.2) -- +(0,.7) (5.5,1.2) -- +(0,.7) (5.5,0) -- +(0,.7);
      \draw[->,thick] (T2) -- +(5.4,0); \draw[->] (T4) -- +(3.5,0); \draw[->] (T5) -- +(.7,0);
      \draw[->,dashed] (5.75,1.55) -- +(1.4,0);
      \draw[->,dashed] (5.75,1.55) to[bend right,->] +(1.2,.7);
      \draw[->,dashed] (5.75,1.55) to[bend left,->] +(1.2,-.7);
    \end{tikzpicture}
    \caption{$T_1$, $T_2$, and $T_4$ create new basic blocks. $T_2$ first reaches the block end and creates new control flow edges. $T_1$ then reaches the block end and is going to split blocks.} \label{fig:inv:sub2}
  \end{subfigure}
  \begin{subfigure}{.9\textwidth}
    \centering
    \begin{tikzpicture}
      \draw (0,1) rectangle +(2,.7) +(2,.35) coordinate(B1e) +(0,.35) node[right]{$B_1$}
          +(0,.7) node[above]{\texttt{0x4}}
        (2.5,1) rectangle +(3.5,.7) +(0,.35) coordinate(B2s) +(0,.35) node[right]{$B_2$}
          +(0,.7) node[above]{\texttt{0xA}}
        (4.2,0) node[below](T5){$T_4$} rectangle +(1.8,.7) +(0,.35) node[right]{$B_3$};
      \draw[->,dashed] (B1e) -- (B2s);
      \draw[dotted] (4.2,1) -- +(0,.7) node[above]{\texttt{0xD}}
        (5.5,1) -- +(0,.7) (5.5,0) -- +(0,.7);
      \draw[->,dashed] (5.75,1.35) -- +(1.4,0);
      \draw[->,dashed] (5.75,1.35) to[bend right,->] +(1.2,.7);
      \draw[->,dashed] (5.75,1.35) to[bend left,->] +(1.2,-.7);
      \draw[->,thick] (T5) -- +(1.3,0);
    \end{tikzpicture}
    \caption{$T_1$ finishes splitting blocks. $T_4$ reaches the block end and is going to split blocks.} \label{fig:inv:sub3}
  \end{subfigure}
  \begin{subfigure}{.9\textwidth}
    \centering
    \begin{tikzpicture}
      \draw (0,0) rectangle +(2,.7) +(2,.35) coordinate(B1e) +(0,.35) node[right]{$B_1$}
          +(0,.7) node[above]{\texttt{0x4}}
        (2.5,0) rectangle +(1.2,.7) +(1.2,.35) coordinate(B2e) +(0,.35) node[right](B2s){$B_2$}
          +(0,.7) node[above]{\texttt{0xA}}
        (4.2,0) rectangle +(1.8,.7) +(0,.35) node[right](B3s){$B_3$}
          +(0,.7) node[above]{\texttt{0xD}};
      \draw[->,dashed] (B1e) -- (B2s); \draw[->,dashed] (B2e) -- (B3s);
      \draw[dotted] (5.5,0) -- +(0,.7);
      \draw[->,dashed] (5.75,.35) -- +(1.4,0);
      \draw[->,dashed] (5.75,.35) to[bend right,->] +(1.2,.7);
      \draw[->,dashed] (5.75,.35) to[bend left,->] +(1.2,-.7);
    \end{tikzpicture}
    \caption{$T_4$ finished splitting blocks, which involves moving edges.} \label{fig:inv:sub4}
  \end{subfigure}
  \caption{An example of five threads work with a common area of code. Solid edges represent the progress of threads. Bold solid edges represent actions to take place. Dashed edges represent control flow edges in the CFG.}
  \label{fig:inv}
\end{figure}

We use \cref{fig:inv} to illustrate how our five invariants ensure that threads correctly perform concurrent CFG operations.

\textbf{Invariant 1: Block Creation.} There is at most one basic block starting at any given address.
This invariant means that if threads branch to the same target concurrently,
one and only one thread should create the block and make the block visible to other threads.
Maintaining this invariant requires efficient concurrent data structures that synchronize threads
branching to the same target, while allowing threads branching to different targets to proceed independently.

In \cref{fig:inv:sub1}, threads $T_3$, $T_4$ and $T_5$ branch to the same address.
According to this invariant, only one thread should create a new basic block.
As shown in \cref{fig:inv:sub2}, $T_4$ creates a new basic block $B_3$.
$T_3$ and $T_5$ do not create any new basic blocks and leave the common code area to work with other code.
Independently, $T_1$ creates basic block $B_1$ and $T_2$ creates basic block $B_2$.

\textbf{Invariant 2: Block End.} There is at most one basic block ending at any given address.
A na\"ive implementaion of this invariant is to let a thread check whether a block exists at its current working address.
If there exists one, the working thread can end its block.
However, this implementation means that there will be a block start lookup after decoding each instruction.
This would create a performance hotspot on the concurrent data structure used for Invariant 1.
Our design defers this check until the working thread reaches a control flow instruction.
In this way, we reduce the frequency of global concurrent data structure lookups from once per instruction to once per control flow instruction.

As shown in \cref{fig:inv:sub2}, thread $T_1$, $T_2$ and $T_4$ will independently parse their blocks
until they reach the indirect jump instruction.
Based on this invariant, only one thread should register the block end address, which is $T_2$ in this example.

\textbf{Invariant 3: Edge Creation.} The thread that registers a block's end is responsible
for creating out-going control flow edges from that block. This invariant ensures that
no redundant control flow edges are created and jump table analysis for a particular indirect jump is always performed by one thread.
This also reduces unnecessary block start lookups by avoiding redundant edges.
As shown in \cref{fig:inv:sub2}, because thread $T_2$ registers the block end, $T_2$ continues to
perform control flow analysis to resolve the indirect jump targets and create control flow edges.
$T_2$ then leaves the common code and continues to work with other code.

\textbf{Invariant 4: Block Split.} The threads that reach a block end but do not register
the block end will need to split blocks. Suppose we have block $B_1[x_1,y), B_2[x_2,y), \ldots B_n[x_n,y)$ created by $n$ threads, where $x_1 < x_2 < \ldots < x_n < y$.
The results of block split should be $B_1[x_1, x_2), B_2[x_2, x_3), \ldots, B_n[x_n, y)$, with a fall-through edge between each pair of adjacent basic blocks.
It is inefficient to wait for all relevant blocks before performing splitting,
so we present the following eager block split algorithm.

Based on Invariant 2 (block end), only one block $B_i[x_i, y)$ will register its end at $y$.
When another block $B_j[x_j, y)$ reaches $y$, the working thread can look up $B_i$ as the registered block. Depending on the relationship between $x_i$ and $x_j$, we have two cases:
\begin{itemize}
\item If $x_i>x_j$, $B_j$ is split into [$x_j$, $x_i$) while $B_i$ is untouched.
We then register $B_j$ at block end address $x_i$, which will trigger a new iteration of block split when another block has already registered block end at $x_i$.
As shown in \cref{fig:inv:sub3}, $T_1$ splits blocks $B_1$, registers $B_1$ ending at $0xA$ and then leaves the common code.

\item If $x_i<x_j$, $B_i$ is split into [$x_i$, $x_j$) while $B_j$ is untouched.
We then replace $B_i$ with $B_j$ for block end address $y$, register $B_i$ for block end address $x_j$, and move out-going edges from $B_i$ to $B_j$.
Similar to the first case, registering $B_i$ at $x_j$ may recursively require another block split.
As shown in \cref{fig:inv:sub4}, $T_4$ splits $B_2$ and moves control flow edges from $B_2$ to $B_3$.
\end{itemize}

For both cases, each iteration of the block split algorithm ends with a smaller block end address. Therefore, our block split algorithm is guaranteed to converge.

\textbf{Invariant 5: Function Creation.} There is at most one function starting at any given address. This invariant has similar properties and requirements to Invariant 1 for creating blocks.

These five invariants ensure that commutative operations can be safely reordered and performed concurrently, and the relative speed of threads will not impact the final results.

\subsection{Parallel Control Flow Traversal}
\label{sec:pp:cft}

\cref{fig:process_a_frame} presents the algorithm for control flow traversal.
Coupled with the invariants defined in \cref{sec:pp:invariants},
control flow traversal can be performed in parallel.

The traversal is repeated until there are no more unanalyzed basic blocks (Line 2).
For each unanalyzed block $b$,
we use routine \inC{linearParsing} to decode instructions until a
control flow transfer instruction is encountered (Line 4).
Modifications to Dyninst's instruction decoding code add thread-safety to support this.

Routine \inC{registerBlockEnd} follows invariant 2 (block end) to register the block end (Line 5).
Only the thread that successfully registers the block end will see a non-empty set of control flow edges returned, following invariant 3 (edge creation).
All other threads reaching the same block end will see an empty set of edges and will follow invariant 4 (block split) to split the blocks (Line 7).

The thread that creates the control flow edges will proceed to traverse the edges (Line 8 - 12).
If we encounter a function call, we may need to create a new function, following invariant 5 (function creation) at Line 10.
If we encounter a call fall-through summary edge or return edge, we process non-returning functions (Line 11).
If we encounter other types of edges, such as indirect, direct or conditional branches,
we create new basic blocks based on invariant 1 (block creation) at line 12.

\textbf{Handling non-returning functions:} In procedure \code{processCall}, we use the non-returning function analysis presented by Meng and Miller~\cite{Meng2016BinaryNotEasy}.
To address the non-returning function dependency between CFG operations,
we improve the analysis to eagerly notify its callers once a function's return status is set to RETURN.
This improvement works well in practice because large functions may contain multiple return instructions.
As soon as we encounter one of a function's return instructions, we know this function is RETURN and we can enable the $O_{CFEC}$ operation in its caller without waiting for analysis of the callee to finish.

\textbf{Jump table analysis.} We address two issues raised in ~\cref{sec:properties} about jump table analysis.
First, jump table analysis ($O_{IEC}$) in Dyninst does not satisfy monotonic ordering property.
We identify that when Dyninst encounters instructions or path conditions that it cannot analyze,
Dyninst will fail to analyze the jump table and generate an empty set of control flow targets.
This issue can be addressed by taking the union of the targets discovered along different paths,
essentially ignoring instructions or path conditions that fail analysis.
In this way, jump table targets identified along valid control flow paths can still be propagated to the indirect jump,
and the analysis can generate non-empty set of control flow targets.
While this strategy makes the jump table analysis in Dyninst satisfies the monotonic ordering property,
it can over-approximate jump table sizes and lead to bogus control flow edges.
We will introduce a cleaning strategy in the CFG finalization stage to remove bogus control flow edges.

Second, the monotonic ordering property specifies that we can get a larger graph if we delay jump table analysis as much as possible,
but this might delay the finding of return instructions and hurt parallelism due to non-returning function dependencies.
We balance these two factors by ordering jump table analysis after the analysis of direct control flow edges in this function,
but before call fall-through edges when the callee does not have a known return status.
In addition, we repeat the analysis of a jump table after more control flow paths are created within the same function.
This fixed-point analysis of jump tables allows us to find most of the targets early in the analysis and gradually converge to a complete set of targets.

\begin{listing}[t]
\begin{minted}[escapeinside=||, mathescape=true]{moon}
  more_func = |$\emptyset$|
  while f.hasMoreBlocks()
    b = f.nextBlock()
    linearParsing(b)
    edges = registerBlockEnd()
    if edges == |$\emptyset$|
      splitBlock(b)
    for e in edges:
      switch e.type()
        when 'call' , 'tailcall' then more_func |$\cup$|= processCall()
        when 'call-fallthrough', 'ret' then more_func |$\cup$|= processNonRetFunc()
        when 'other': createNewBasicBlock(f)
  return more_func
\end{minted}
\caption{\small The algorithm of control flow traversal. }
\label{fig:process_a_frame}
\end{listing}

\subsection{CFG Finalization} \label{sec:pp:finalization}

The goal of the CFG finalization stage is remove wrong CFG elements and determine function boundaries. No new CFG elements will be added.

The first step is jump table finalization, where we remove wrong indirect control flow edges.
We find that over-approximation of jump targets is caused primarily by over-approximation of jump table sizes.
We can mitigate this problem by leveraging an observation that compilers do not emit overlapping jump tables~\cite{WilliamsKing2020Egalito}.
Therefore, if the analysis of a jump table overflows into another jump table, we can detect over-approximation
and apply edge removal operations $O_{ER}$ to remove wrong edges and cascading dangling blocks.
We make two observations about this strategy.
First, we have established in~\cref{sec:properties} that edge removal operations commutate.
Therefore, it is safe to perform this mitigation strategy in parallel.
Second, this strategy cannot be used during the parallel control flow traversal step.
This is because when we analyze a jump table, we do not know the exact locations of all jump tables in the binary.
For this reason, we delay this mitigation of over-approximation until the CFG finalization phase.

We then address wrong tail calls edges and determine function boundaries.
We handle this with an iterative parallel control flow graph search.
Starting from function entries, we add blocks to the boundary of a function
by traversing intra-procedural edges.
After getting the temporary function boundaries, we use three rules in order to correct tail call results:
\begin{enumerate}
\item If a branch is marked as not a tail call, but the edge target has a CALL incoming edge, we correct this edge to be a tail call.
\item If a branch is marked as a tail call, but the branch target is within the current function boundary, we correct this edge to be not a tail call.
\item If a branch is currently a tail call, but the edge target has only the current edge as incoming edges, we treat this as not a tail call. This is generally caused by outlined code blocks.
\end{enumerate}

After correcting tail calls, we re-perform the function boundary graph search and the tail call correction procedure.
We flip the determination of tail call at most once for each edge, ensuring convergence.

Finally, we remove functions that do not have incoming inter-procedural edges.
\section{Implementation Experiences}
\label{sec:implementation}

We implemented our new parallel CFG construction algorithms in Dyninst.
Careful implementation that follows our design is crucial for correctness and high performance.
We present several code examples and lessons we learned in our work.

\subsection{Sample Implementations of Invariants} \label{ssec:tbb}
In \cref{sec:pp:invariants} we presented five invariants for parallel control flow traversal.
An efficient implementation of these is the foundation for scalable parallel binary code analysis.

\cref{lst:invariant:createblock} is a code example of our implementation for invariant 1 (block creation).
This code example can be easily adapted to implement invariant 5 (function creation).
Recall that two requirements for invariant 1 are
(1) threads that branch to the same address should be synchronized and only one thread should create a new block, and
(2) threads that branch to different addresses can make progress independently.
Our implementation uses the concurrent hash map provided by Intel's Threaded Building Blocks library~\cite{tbb} to fulfill these two requirements,
which provides entry-level reader-writer locks.
The \inC{insert} method of \inC{concurrent_hash_map} ensures that only one of the concurrent insertions with the same key will succeed (Line 5).
Therefore, we can use the return value of \inC{insert} to determine whether the current thread has successfully created a block and should continue analysis of the block (Line 7).
Threads that see a \inC{false} return value knows that another thread has created the block and can move on to other work (Line 9 - 10).

\begin{listing}[t]
\begin{minted}{cpp}
  tbb::concurrent_hash_map<Address, Block*> blocks;
  bool attemptToCreateBlock(Address a) {
    Block* b = new Block(a);
    if (blocks.insert({a, b})) {
      // Successfully registered the new block.
      return true;
    } else { // Block already exists.
      delete b;
      return false;
    }
  }
\end{minted}
\caption{Example implementation of invariant 1 (block creation).}
\label{lst:invariant:createblock}
\end{listing}

\cref{lst:invariant:splitblock} is a code example showing how invariant 2 (block end), invariant 3 (edge creation), and invariant 4 (block split) fit together.
\inC{concurrent_hash_map} exposes the entry-level reader-writer locks via an ``accessor'' semantic.
We can obtain an ``accessor'' for the existing entry in the table (inserting one if requested and not already present).
The accessor acts as a read or write lock on the entry, and other threads that are trying to obtain a conflicting accessor will wait until the holding thread releases its own accessor.
Line 4 ensures only one block is registered for a block end address, enforcing invariant 2.
The accessor ensures that edge creation (Line 6) and block splitting (Line 10) are mutually exclusive.
This mutual exclusion guarantees that control flow edges will not be created while being moved.
Note that invariant 3 and 4 do not require mutual exclusion,
and it is possible to implement them with finer grained synchronization.
Our implementation uses mutual exclusion for simplicity and our performance profiling has not shown this mutual exclusion to be a performance bottleneck.

\begin{listing}[t]
\begin{minted}{cpp}
  tbb::concurrent_hash_map<Address, Block*> blocks_end;
  bool blockEnd(Block* b) {
    tbb::concurrent_hash_map<Address, Block*>::accessor a;
    if (blocks_end.insert(a, b->end())) {
      // Block end registered, continue to create edges.
      AnalyzeCFEdges(b);
      return true;
    } else {
      // a->second references the block in the entry.
      a->second = splitBlock(b, a->second);
      return false;
    }
  }
\end{minted}
\caption{Example implementation for invariant 2 (block end), invariant 3 (edge creation), and invariant 4 (block split).}
\label{lst:invariant:splitblock}
\end{listing}

\subsection{Multi-Keyed Parallel Symbol Table} \label{ssec:indexed_symbols}

In~\cref{sec:pp:stages}, we described that it is necessary to have a parallel symbol table as a large binary may contain millions of symbols.
Dyninst's symbol table supports lookups by any of its four properties: byte offset, mangled name, ``pretty'' human-readable name
and demangled ``typed'' name.
The original implementation used a template class from Boost~\cite{boost} to implement this, a
very customizable structure called a \inC{multi_index_container}.
Since the Boost implementation is not thread-safe, after contention for its mutex lock
became a notable bottleneck we redesigned
the structure for concurrency as shown in \cref{lst:multikeymap}.

The key insight is that no lookups occur in parallel with an insertion or modification, so synchronization
is only needed during writes.
Two writes only conflict if the \inC{Symbol} they are working with is the same, so we use the
entry-level lock on the \inC{master} table to mediate between threads.
The thread which inserts on the \inC{master} table proceeds to update the corresponding entries in
the \inC{by*} tables, retaining its lock to ensure that any other modifications to the collective
entries occur in a total order.
Once all modifications are complete, later lookups are able to use the \inC{by*} tables directly,
giving the same semantics as the original structure.

\begin{listing}[t]
  \begin{minted}{cpp}
    class Symtab::indexed_symbols {
      concurrent_hash_map<Symbol*, Offset> master;
      concurrent_hash_map<Offset, vector<Symbol*>> byOffset;
      concurrent_hash_map<std::string, vector<Symbol*>> byMangledName;
      concurrent_hash_map<std::string, vector<Symbol*>> byPrettyName;
      concurrent_hash_map<std::string, vector<Symbol*>> byTypedName;
      bool insert(Symbol* s) {
        accessor a;
        if(!master.insert(a, {s, s->getOffset()}))
          return false;  // Already inserted, no need to continue
        { accessor a1;
          byOffset.insert(a1, s->getOffset());
          a1->second.push_back(s); }
        { accessor a2;
          byMangledName.insert(a2, s->getMangledName());
          a2->second.push_back(s); }
        // ... etc. ...
        return true;
      }
    };
  \end{minted}
  \caption{Implementation example for a thread-safe efficient map with multiple keys,
  discussed in \cref{ssec:indexed_symbols}.}
  \label{lst:multikeymap}
\end{listing}

\subsection{Performance Improvements}

We summarize two implementation lessons that are beneficial for improving performance,
starting with replacing parallel loops with task parallelism.
As described in \cref{sec:parseapi}, we use a parallel for loop to perform parallel control flow traversal
and collect new functions to analyze.
The problem with this implementation is that analysis of newly collected functions will not start until
all existing functions have been analyzed.
This can cause significant idleness when the analysis of functions is imbalanced.
To address this issue, our improved implementation uses OpenMP tasks as the parallel programming model
and we launch a new task as soon as we discover a new function to analyze.

The second lesson is to use a thread local cache to reduce redundant calculations while not incurring thread synchronization overheads.
For invariant 2 (block end) discussed in \cref{sec:pp:invariants},
we let each thread parse their blocks without any synchronization until reaching a control flow instruction.
This design causes redundant instruction decoding between overlapping blocks analyzed by different threads.
However, while functions sharing code is common, most of the code blocks in a binary are still not shared.
This means that most of the time, a thread is going to branch into a block that was created by itself, not created by other threads.
Therefore, we implemented a thread local cache that maintains the addresses that have been analyzed by the thread
and use this cache to reduce redundant decoding.

\subsection{Profiling and Debugging Tools} \label{sec:implementation:tool}

The implementation of our new parallel CFG construction algorithms in Dyninst
involves a large amount of code, over 120K lines of code related to reading ELF sections, decoding machine instructions, and data flow analysis.
We find that the following workflow focuses our attention to what needed it most, and helps us identify numerous thread-safety issues across the code base.

\begin{enumerate}
\item Use a performance analysis tool such as HPCToolkit~\cite{HPCToolkit2010} to gather performance traces of the code
with the aim of identifying code regions whose computational cost justifies an overhaul to add parallelism.

\item Inspect the identified code and choose a candidate parallelization strategy,
such as loop-level parallelism, a fork-join parallel programming
model, or task-level parallelism.
The right choice for a particular code region depends on its code
structure as well as the algorithms and data structures that it
employs.

\item Use data race detectors to identify data structures and sharing patterns that require synchronization.
Candidate tools include logical data race detectors such as \texttt{cilkscreen} and
happens-before race detectors such as \texttt{helgrind}, the latter available as part of the Valgrind binary instrumentation system~\cite{ValgrindGeneral}.

\item Use the performance analysis tool to identify excessive mutual exclusion, unbalanced workloads,
and excessive phase-based synchronization that form major bottlenecks.

\item Test the design and implementation with a large suite of test cases and benchmarks.
\end{enumerate}

The above steps are repeated until the overall performance improves satisfactorily.

\section{Application Case Studies} \label{sec:application}

We present two application case studies that utilize our new parallel CFG construction algorithms:
\texttt{hpcstruct}, a utility in HPCToolkit for performance analysis,
and \texttt{BinFeat}, a feature extraction tool for software forensics.

\subsection{Application Background}

Besides constructing CFGs (AC1),
other commonly used analysis capabilities by binary analysis applications include
(AC2) identifying loops,
(AC3) building a mapping between source lines and machine instructions,
(AC4) understanding function inlining for templates and inlined functions,
(AC5) iterating over functions, basic blocks, edges, and machine instructions, and
(AC6) performing data flow analysis such as register liveness analysis.
We use two application examples to illustrate how these analysis capabilities are used.

\textbf{Performance Analysis with HPCToolkit}:
HPCToolkit is an integrated suite of tools for measurement and analysis of
application performance on computers ranging from desktops to supercomputers.
To relate performance measurements to an application's source code,
the \texttt{hpcstruct} utility in HPCToolkit relates each
machine instruction address to the static calling context in which it occurs.
In particular, \texttt{hpcstruct} is able to relate instructions to their original function (AC1) or loop construct (AC2)
by inspection of the binary's final CFG (AC5), and to an inlined function or template (AC4) and source lines (AC3) if
DWARF debugging information is available.

\textbf{Feature Extraction with BinFeat}:
BinFeat is a tool for extracting binary code features for software forensics tasks,
including function entry identification, compiler identification and authorship attribution.
Commonly used features include machine instruction sequences, subgraphs of CFGs (AC1),
loop nesting levels (AC2), and live register counts (AC6).
BinFeat iterates over all functions and blocks to extract these features (AC5).

\subsection{Application Parallelization}
Even if Dyninst's CFG construction algorithms are parallelized,
binary analysis applications need to reduce serial execution to achieve good speedup.
The basic idea here is that after the CFG has been fully constructed, binary analysis will typically no longer make modifications to the CFG.
Therefore, the CFG becomes read-only and different threads can safely perform analysis independently as long as the analysis itself is thread-safe.
Based on this idea, we summarize a design pattern for parallelizing binary analysis applications.

\begin{listing}[t]
  \begin{minted}{cpp}
    ParseAPI::CodeObject *co = getCodeObject();
    co->parse() // Perform CFG construction in parallel
    std::vector<ParseAPI::Function*> funcs = co->funcs();
    SortFuncs(funcs); // Sort functions to address load balancing
    // Parallel for loop to analyze each function in parallel
    #pragma omp parallel for schedule(dynamic)
    for (size_t i = 0; i < fvec.size(); ++i) {
      ParseAPI::Function* f = fvec[i];
      ParseAPI::LoopAnalyzer la(f); // Analyze loops
      DataflowAPI::LivenessAnalyzer live(f); // Register liveness analysis
      DataflowAPI::StackAnalysis sa(f); // Stack height analysis
      // Other thread-safe intra-procedural analysis
    }
  \end{minted}
  \caption{Code example of utilizing Dyninst for parallel binary analysis applications.}
  \label{lst:app:temp}
\end{listing}

\cref{lst:app:temp} shows an example code snippet to write parallel binary analysis applications.
Line 2 uses the parallel CFG construction algorithm described in \cref{sec:parseapi} to construct a CFG.
Line 3 and 4 get the list of functions in the binary and sort the functions to address load balancing between threads.
Sorting is important as functions will have different sizes, which can cause notable unbalance if
a large function is scheduled last in a work queue.
Therefore, we sort the functions in decreasing order so that large functions are processed first.
Within the parallel loop, the user can apply intra-procedural analysis in parallel to different functions.

To complete the parallelization of a binary analysis application, an application developer will also need to parallelize
application-specific logic.

For \texttt{hpcstruct}, we parallelize the parsing of DWARF debugging information in a binary.
A binary's DWARF information is organized in a forest-like
structure with a tree for each compilation unit (CU).
Since source files are typically of similar sizes across a project % (in the absence of generated code),
we simply used an OpenMP \inC{parallel for} loop to process each of the CUs in parallel, accumulating
their information in structures allocated in parallel by a previous phase.
This resulted in thousands of race reports, which we handled first by mutex locks and
then later by using concurrent data structures such as those discussed in \cref{ssec:tbb}.
Some races were caused by code within Libdw, a utility library from Red Hat for parsing DWARF, and in cases where the performance would suffer from full
mutual exclusion we applied more significant modifications by implementing a resizeable hash table~\cite{michael2002high,click2007lock,triplett2011resizable}
in Libdw.

BinFeat needs to build a global feature index after extracting features from every functions in a binary.
This operation can be parallelized with a reduction operation, which is a generic parallel computing primitive.

\section{Evaluation} \label{sec:evaluation}

Since it is a challenging task to generate accurate ground truth for a binary's CFG,
we evaluate the correctness of our parallel CFG construction algorithm and implementation
by approximating the ground truth with debug information and RTL intermeidates.
We then evaluate the performance of our work using \texttt{hpcstruct} and BinFeat.

\subsection{Correctness}
To illustrate the correctness of our approach, we verified our algorithm and implementation using 113 binaries obtained by compiling the coreutils and tar projects.
These binaries are compiled with GCC 9.3.0 for x86-64, with
link-time optimization disabled and other optimizations enabled as specified by the package.
In addition, we compiled these binaries with debug information and injected the flag \texttt{-fdump-rtl-dfinish} to generate RTL intermediates for individual source files.
The debug information and RTL are used only for generating the ground truth.

The ground truth of this data set consists of three parts:

\begin{itemize}
\item We represent the boundary of function with address ranges, essentially projecting the CFG of a function to the virtual address space.
The DWARF \code{.debug\_info} section encodes function ranges. In particular, it supports multiple non-contiguous ranges for one function and supports one range corresponding to multiple functions.
Therefore, we can evaluate the handling of functions sharing code and non-contiguous functions.
\item We include the size of a jump table as part of the ground truth, which can be extracted by scanning the RTL files.
Unfortunately, we cannot derive jump table locations or the actual targets from the RTL files.
As existing jump table analysis has focused on bounding the size of jump tables, we believe jump table sizes provide significant evaluation value.
\item RTL encodes the ground truth for calls to non-returning functions, where a non-returning call has \code{REG\_NORETURN} as one of its arguments.
\end{itemize}

We then write a checker program that uses our parallel CFG construction implementation to get the CFG,
print out function ranges, jump table sizes, and non-returning calls,
and match these items with the ground truth.

We identified four distinct differences between our implementation and the ground truth by manual inspection of the automatically identified differences:

\begin{itemize}
\item Failing to identify non-returning calls to `\texttt{error}', causing functions to include additional ranges.
`\texttt{error}' is non-returning when the first argument is non-zero, but returning when the first argument is zero.
Existing non-returning function analysis performs name matches for external functions. This approach does not work for `\texttt{error}'.
\item For a function \texttt{foo}, the compiler may emit another function symbol (``\texttt{foo.cold}'') for outlined cold blocks from \texttt{foo}.
However, the debugging information does not encode ``\texttt{foo.cold}'' and lists the address ranges of ``\texttt{foo.cold}'' blocks as part of \texttt{foo}.
\item Failing to resolve a jump table whose calculation uses the stack to store intermediate values.
\item An extra indirect jump target caused by failing to identify a non-returning call to `\texttt{error}', leading bogus wrong control flow edge to the indirect jump.
\end{itemize}

In all cases above, the differences are caused by either incorrectness in the individual CFG operations ($O_{CFEG}$ and $O_{IEC}$)
or mismatches between the symbol table and the DWARF information.
In other words, the errors are not caused incorrect parallelism and can be fixed by improving the implementations of $O_{CFEG}$ and $O_{IEC}$.

\subsection{HPCToolkit's \texttt{hpcstruct}}
\label{sec:eval:stages}
\newcommand\ares{LLNL1}
\newcommand\kull{LLNL2}
We use four large binaries to illustrate the effectiveness of our parallelization for speeding up performance analysis,
including two binaries from Lawrence Livermore National Laboratory (\ares{} and \kull{}\footnote{
  Due to export control, we are unable to disclose the names of these binaries until approved by LLNL.
  }), one large binary from Argonne National Laboratory (Camellia),
  and one shared library from TensorFlow~\cite{tensorflow}.

Sizes of relevant sections of the four binaries are given in \cref{tab:binaries}.
\ares{} is a Power little-endian 64-bit binary, \kull{} and TensorFlow are x86-64,
and Camellia is a Power big-endian 32-bit binary.
\ares{}, \kull{} and Camellia were compiled by their corresponding software development teams,
we compiled the TensorFlow binary with GCC 8.3.0.
Experiments run on LLNL binaries were run on a node with 16 threads (8 cores), Camellia on one with
36 threads (18 cores), and TensorFlow on a two-socket machine with 36 cores each (72 threads total).

\begin{table}[t]
\small
  \setlength{\tabcolsep}{3pt}
  \sisetup{
    table-format = 4.2,
    table-text-alignment = right,
  }
  \begin{tabular}{rSSS} \toprule
    & \multicolumn{3}{c}{Section(s) Sizes (MiB)} \\ \cmidrule(l){2-4}
    Binary     & {Total} & \texttt{.text} & \texttt{.debug\_*} \\ \midrule
    \ares      &  363.40 &    77.01       &   243.16           \\
    \kull      & 1913.50 &   149.13       &  1612.20           \\
    Camellia   &  299.08 &    40.81       &   232.43           \\
    TensorFlow & 7844.81 &   112.21       &  7622.46           \\ \bottomrule
  \end{tabular}
  \caption{Relevant statistics of the binaries used as input for the various benchmarks.}
  \label{tab:binaries}
\end{table}

\begin{figure}[t]  
  \adjincludegraphics[width=.48\textwidth,frame=0.5px]{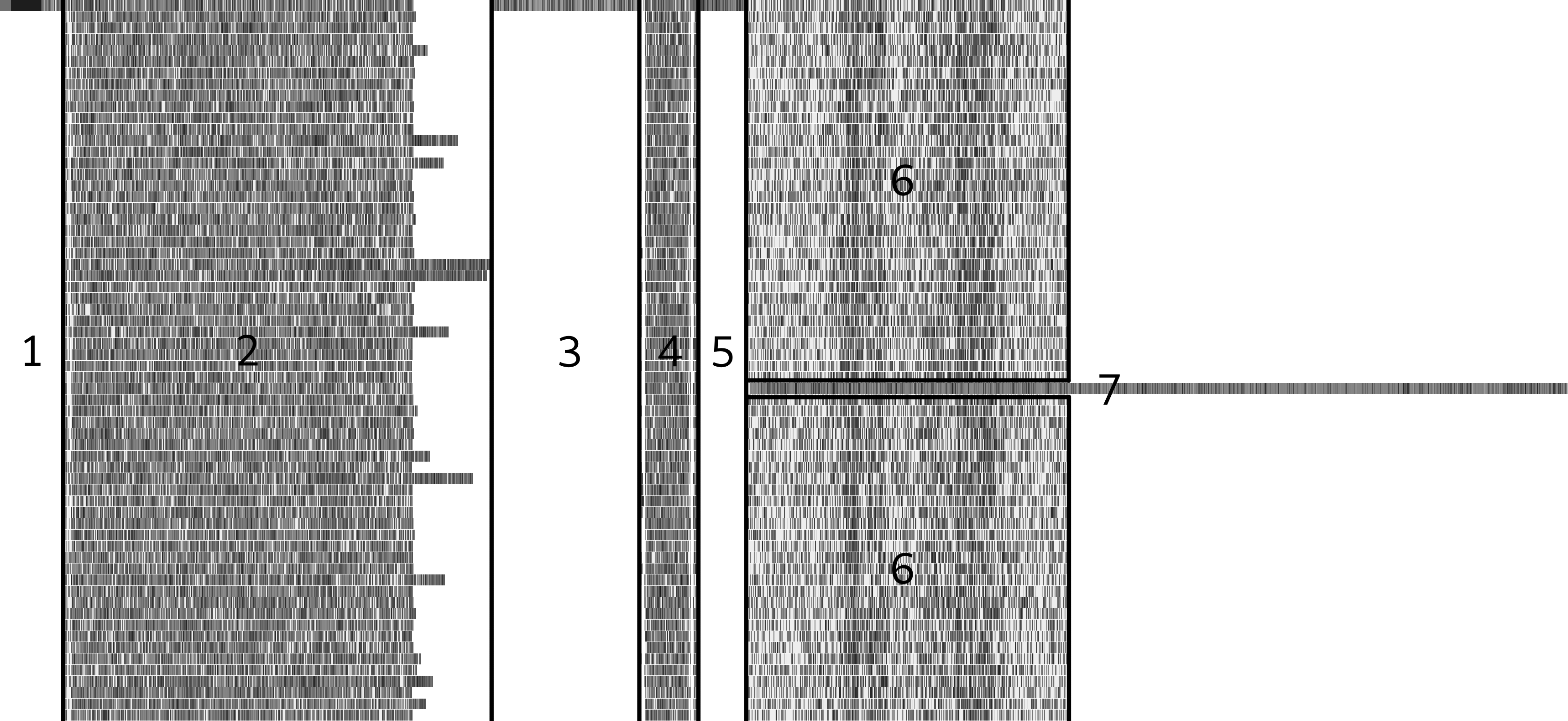}
  \centering
  \caption{Trace from a run of \texttt{hpcstruct} on TensorFlow, descriptions of labeled sections are
    given in \cref{sec:eval:stages}.}
  \label{fig:structtrace}
\end{figure}

\begin{table}[t]
\small
\sisetup{
  table-number-alignment = left,
  table-format = 3.2,
}
\newcommand{\xelem}[1]{\multicolumn{1}{S[
  table-space-text-pre = {$\times$\,},
  table-format = 2.2,
]}{{$\times$\,} #1}}
\setlength{\tabcolsep}{3pt}
\begin{tabular}{lrSSS[separate-uncertainty,table-format=3.2(3)]} \toprule
\multicolumn{2}{l}{Binary}  & \multicolumn{3}{c}{Time Taken (s)} \\ \cmidrule(l){3-5}
\multicolumn{2}{r}{Cores}   & {\inC{DWARF} (2)}  & {\inC{CFG} (4)}   & \inC{hpcstruct} \\
\midrule \multirow{3}{*}{\ares}
                       &  1 &         30.44 &        101.57 & 237.97 +- 3.79  \\
                       \addlinespace[-1pt]
                       % &  2 &         15.77 &         56.10 & 131.62 +- 1.19  \\
                       % \addlinespace[-1pt]
                       % &  4 &          8.22 &         31.29 &  74.76 +- 1.42  \\
                       % \addlinespace[-1pt]
                       % &  8 &          4.44 &         17.81 &  44.17 +- 0.33  \\
                       % \addlinespace[-1pt]
                       & 16 &          2.65 &         11.21 &  30.44 +- 0.28  \\
\multicolumn{2}{r}{Speedup} & \xelem{11.47} & \xelem{ 9.06} & \xelem{ 7.82}   \\
                       \addlinespace[-1pt]
\midrule \multirow{3}{*}{\kull\footnotemark}
                       &  1 &         83.95 &        176.79 & 690.86          \\
                       \addlinespace[-1pt]
                       % &  2 &         40.00 &         92.96 & 386.61          \\
                       % \addlinespace[-1pt]
                       % &  4 &         20.50 &         49.97 & 213.01          \\
                       % \addlinespace[-1pt]
                       % &  8 &         10.67 &         27.65 & 130.03          \\
                       % \addlinespace[-1pt]
                       & 16 &          6.07 &         19.66 & 112.55          \\
\multicolumn{2}{r}{Speedup} & \xelem{13.83} & \xelem{ 8.99} & \xelem{ 6.14}   \\
                       \addlinespace[-1pt]
\midrule \multirow{3}{*}{Camellia~\cite{camellia}}
                       &  1 &         22.36 &         46.10 & 118.39 +- 2.24  \\
                       \addlinespace[-1pt]
                       % &  2 &         12.18 &         27.18 &  67.93 +- 1.51  \\
                       % \addlinespace[-1pt]
                       % &  4 &          7.11 &         15.22 &  38.96 +- 0.96  \\
                       % \addlinespace[-1pt]
                       % &  8 &          4.67 &          8.61 &  24.63 +- 0.56  \\
                       % \addlinespace[-1pt]
                       & 16 &          3.34 &          5.38 &  20.21 +- 0.17  \\
\multicolumn{2}{r}{Spd.}    & \xelem{ 7.86} & \xelem{11.42} & \xelem{ 5.86}   \\
\midrule \multirow{4}{*}{TensorFlow~\cite{tensorflow}}
                       &  1 &          702.81 &          112.55 & 1252.88 +- 19.67  \\
                       \addlinespace[-1pt]
                       % &  2 &        374.51 &              59.18 &  672.00 +- 3.86  \\
                       % \addlinespace[-1pt]
                       % &  4 &        200.09 &              31.47 &  368.73 +- 5.21  \\
                       % \addlinespace[-1pt]
                       % &  8 &        109.68 &              17.34 &  217.08 +- 2.38  \\
                       % \addlinespace[-1pt]
                       & 16 &        64.29 &              9.56 &  160.82 +- 3.08  \\
                       \addlinespace[-1pt]
                       & 32 &        49.63 &              5.49 &  146.12 +- 1.70  \\
                       \addlinespace[-1pt]
                       & 64 &        48.67 &              4.46 &  154.61 +- 2.86  \\
\multicolumn{2}{r}{Spd.}    & \xelem{14.44} & \xelem{25.22} & \xelem{ 8.103}   \\
\bottomrule
\end{tabular}
\caption{Performance results, averages of 10 runs unless otherwise noted.
Times for \texttt{DWARF} and \texttt{CFG} represent parallel DWARF parsing
and parallel CFG construction, corresponding to sections 2 and 4 in \cref{fig:structtrace}.}
\label{tab:perf}
\end{table}

\pgfplotstableread{
  t  ares   aressp kull   kullsp camellia camsp tf      tfsp  gm    min
  1  237.97 1      690.86 1      118.39   1     1252.88 1     1     1
  2  131.62 1.808  386.61 1.787  67.93    1.743 672.00  1.864 1.800 1.743
  4  74.76  3.183  213.01 3.243  38.96    3.039 368.73  3.398 3.213 3.039
  8  44.17  5.388  130.03 5.313  24.63    4.806 217.08  5.772 5.308 4.806
  16 30.44  7.818  112.55 6.138  20.21    5.858 160.82  7.791 6.841 5.858
  32 -      -      -      -      -        -     146.12  8.574 8.574 8.574
  64 -      -      -      -      -        -     154.61  8.103 8.103 8.103
}{\structperf}
\pgfplotstableread{
  t  ares   aressp  kull   kullsp camellia camsp tf     tfsp   gm     min
  1  30.44  1       83.95  1      22.36    1     702.81 1      1      1
  2  15.77  1.930   40.00  2.099  12.18    1.836 374.51 1.877  1.933  1.836
  4  8.22   3.703   20.50  4.095  7.11     3.145 200.09 3.512  3.597  3.145
  8  4.44   6.856   10.67  7.868  4.67     4.788 109.68 6.408  6.378  4.788
  16 2.65   11.487  6.07   13.83  3.34     6.695 64.29  10.931 10.384 6.695
  32 -      -       -      -      -        -     49.63  14.161 14.161 14.161
  64 -      -       -      -      -        -     48.67  14.441 14.441 14.441
}{\symtabperf}
\pgfplotstableread{
  t  ares   aressp kull   kullsp camellia camsp tf     tfsp   gm     min
  1  101.57 1      176.79 1      46.10    1     112.55 1      1      1
  2  56.10  1.811  92.96  1.902  27.18    1.696 59.18  1.902  1.826  1.696
  4  31.29  3.246  49.97  3.538  15.22    3.029 31.47  3.576  3.400  3.029
  8  17.81  5.703  27.65  6.394  8.61     5.354 17.34  6.492  5.967  5.354
  16 11.21  9.061  19.66  8.992  5.38     8.569 9.56   11.767 9.520  8.569
  32 -      -      -      -      -        -     5.49   20.490 20.490 20.490
  64 -      -      -      -      -        -     4.46   25.222 25.222 25.222
}{\parseperf}

\begin{figure}[t]
\begin{tikzpicture}
\begin{loglogaxis}[
  xlabel={Threads}, ylabel={Speedup}, log basis x=2, log basis y=2,
  error bars/y dir=both, error bars/y explicit,
  no markers, legend pos=north west,
  log ticks with fixed point,
]
\addplot+ table [x=t, y expr={\thisrow{gm}},
  y error plus expr={max(\thisrow{aressp},\thisrow{kullsp},\thisrow{camsp},\thisrow{tfsp}) - \thisrow{gm}},
  y error minus expr={\thisrow{gm} - \thisrow{min}},
] {\structperf};
\addlegendentry{\texttt{hpcstruct}};
\addplot+ table [x=t, y expr={\thisrow{gm}},
  y error plus expr={max(\thisrow{aressp},\thisrow{kullsp},\thisrow{camsp},\thisrow{tfsp}) - \thisrow{gm}},
  y error minus expr={\thisrow{gm} - \thisrow{min}},
] {\symtabperf};
\addlegendentry{\texttt{DWARF}};
\addplot+ table [x=t, y expr={\thisrow{gm}},
  y error plus expr={max(\thisrow{aressp},\thisrow{kullsp},\thisrow{camsp},\thisrow{tfsp}) - \thisrow{gm}},
  y error minus expr={\thisrow{gm} - \thisrow{min}},
] {\parseperf};
\addlegendentry{\texttt{CFG}};
\end{loglogaxis}
\end{tikzpicture}
\caption{Average speedup (geometric mean) of \texttt{hpcstruct} on the four binaries, as described in
\cref{sec:eval:stages}.}
\label{fig:perf}
\end{figure}
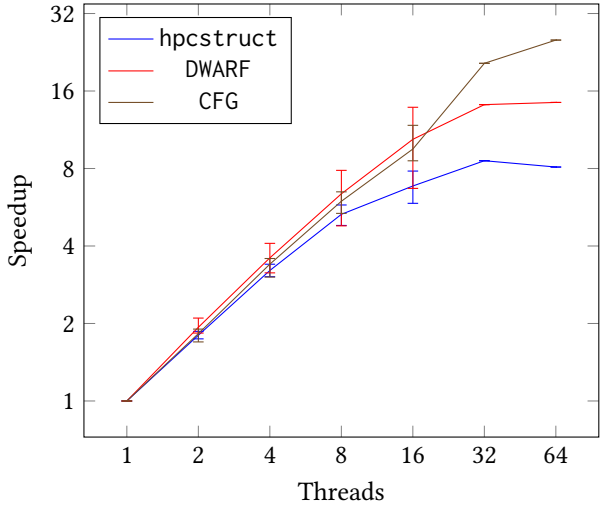

\footnotetext{Results for \kull{} are based on one run for each thread count. We have limited
access to the binary and cannot repeat the experiment.}

The results are presented in \cref{tab:perf} and in \cref{fig:perf}.
Overall, \texttt{hpcstruct} has an end-to-end speedup of $6\times$ to $8\times$, due to several serial phases in
the application code.
We achieved a speedup of $9\times$ to $25\times$ for constructing CFGs
and a speedup of $8\times$ to $14\times$ for DWARF parsing.

To better understand the end-to-end performance impact of our work,
we break down the main phases of execution within \texttt{hpcstruct} in \cref{fig:structtrace},
which presents a performance trace of \texttt{hpcstruct} running on TensorFlow with 64 threads.
The contents of each phase are as follows:
\begin{enumerate}
\item Read data from disk into an internal buffer.
\item Parse DWARF type information in parallel and store in appropriate data structures.
      Imbalance in the sizes of compilation units can cause some idling.
\item Parse address to function and line mappings from DWARF and store in a serial structure
      optimized for accelerated lookup.\footnote{The design of the
        data structure used here makes this region difficult to
        parallelize.}
\item Parse text regions in parallel to identify functions and construct the final
      CFG.
      % Our main work resides in this region.
\item Convert line map and parsing results into ``skeleton'' objects inside \texttt{hpcstruct}, which are suitable for export.
\item Query Dyninst structures in parallel to fill the ``skeleton'' with the final data to
      be serialized.
\item Serialize data and write to disk in parallel with queries to mitigate the effects of serial
      processing.
\end{enumerate}

Although our parallelization (2 and 4) scales well, the overall execution of \texttt{hpcstruct}
has difficulties scaling.
As per Amdahl's Law, the serialization in application code (1, 5-7) and remaining
difficulties (3) prevent our speedup from scaling past $13\times$.
Applications with less serialization will see larger speedups.

\subsection{\texttt{BinFeat}}

Software forensic researchers typically use real world software to construct their training sets.
We follow their practice and construct a set of binaries to analyze.
We compiled Apache HTTP Server~\cite{httpd}, Redis~\cite{redis}, Mysqlslap~\cite{mysqlslap}, and Nginx~\cite{nginx} with GCC-6.4.0 and -O2 optimization.
Our data set contains 504 binaries.
This experiment was run on a x86-64 machine with 18 cores, 72 threads, and 48MB of L3 cache.

\begin{table}[t]
  \small
  \sisetup{
    table-number-alignment = left,
    table-format = 3.2,
  }
  \newcommand{\xelem}[1]{\multicolumn{1}{S[
    table-space-text-pre = {$\times$\,},
    table-format = 2.2,
  ]}{{$\times$\,} #1}}
  \setlength{\tabcolsep}{3pt}
  \begin{tabular}{rSSSSS} \toprule
  \multicolumn{5}{c}{Time Taken (s)} \\ \cmidrule(l){2-6}
  Cores   & \inC{CFG}    & \inC{IF}  & \inC{CF}   & \inC{DF}  & \inC{BinFeat}\\
                           1 &          231.90 &         246.33 &        108.46 &      307.88  & 915.36  \\
                         \addlinespace[-1pt]
                           2 &          142.15 &         125.29 &        56.06 &       173.16  & 518.06  \\
                         \addlinespace[-1pt]
                           4 &          96.95 &          66.56 &         29.54 &        99.02  & 312.48  \\
                         \addlinespace[-1pt]
                           8 &          75.92 &          36.64 &         16.41 &        62.91  & 211.77  \\
                         \addlinespace[-1pt]
                          16 &          64.40 &          22.35 &         9.76 &         44.88  & 160.76  \\
                         \addlinespace[-1pt]
                          32 &          58.47 &          14.37 &         6.27 &         34.62  & 130.43  \\
                         \addlinespace[-1pt]
                          64 &          60.40 &          13.80 &         6.93 &         34.23  & 131.90  \\
  Speedup & \xelem{ 3.84} & \xelem{17.85} & \xelem{ 15.66} & \xelem{ 9.00}  & \xelem{ 6.94} \\
                         \addlinespace[-1pt]
  \bottomrule
  \end{tabular}
  \caption{Performance results for \texttt{BinFeat}. \texttt{CFG}, \texttt{IF}, \texttt{CF}, \texttt{DF} represent the stages of
  CFG construction, extracting instruction features, control flow features, and data flow features, respectively.}
  \label{tab:perf:BinFeat}
  \end{table}

\cref{tab:perf:BinFeat} shows the performance results for \texttt{BinFeat}.
We achieved 7$\times$ overall speedup using 32 hardware threads, but did not gain any further improvement with 64 threads.
Extracting instruction features (18$\times$) and control flow features (16$\times$) scale well to 64 threads.

The extraction of data flow features only achieves 9$\times$ maximium speedup,
we find that its performance is hurt by imbalanced workload between threads.
Note that we extract features from each function in parallel.
Data flow analysis typically has a higher time complexity compared to analyzing instructions and traversing control flow graphs.
Therefore, the analysis of large functions will dominate the whole execution.

CFG construction has only 4$\times$ speedup,
we identify two factors that limit its performance.
First, the issue of imbalanced workload also applies to CFG construction as
the jump table analysis in CFG construction takes significantly longer to run compared to other CFG operations such as creating direct edges.
Second, as described in~\cref{sec:properties}, the non-returning function dependencies between CFG operations can hurt parallelism.
While we mitigate this problem with an eager approach discussed in~\cref{sec:pp:cft},
this problem still persists.
Note that these two issues do not show up for large binaries such as those used in the \texttt{hpcstruct} experiments.
We find that large binaries contain sufficient numbers of functions to keep threads busy and hide these two issues.

\section{Discussion}

\textbf{Benefiting other applications:}
Our work provides a general framework for researchers to parallelize their binary analysis applications.
For example, software vulnerability searching calculates binary code similarity~\cite{david2016statistical,Chandramohan2018BinGo} to match known vulnerable code.
The calculation of binary code similarity utilizes binary analysis capabilities of analyzing machine instruction characteristics, control flow, and data flow.
Our work has parallelized several common analysis capabilities and it will be interesting to see how our work benefits other binary analysis applications.

\textbf{Compiler assisted analysis:}
Our work opportunistically uses information from the compiler (such as providing correct and detailed labels in the code and DWARF).
However, this is not a complete solution and we cannot rely on sufficient or even accurate compiler support.
Surprisingly often for even the most widely-used compilers, the compiler-provided information is incomplete or inaccurate.
One key issue is that binary analysis applications do not typically control which compiler is used to generate the input binaries.
For performance analysis, software developers often use the compiler and optimization flags that lead to greatest performance,
which often leads to less accurate debugging information.
Software forensic analysts deal with binaries collected from the wild, whose compiler generated information is often intentionally removed to
defend against analysis.
Therefore, while we use compiler assistance when available, we cannot not rely on its presence.

\textbf{Other forms of parallelism:}
We focus on multi-threading as the mechanism for parallelization.
Other forms of parallelism can be used to further improve the performance of binary analysis applications.
For example, \texttt{BinFeat} can benefit from node level parallelism by distributing the analysis of different binaries
to different machines.
We believe this type of parallelism is possible for certain specific applications,
and is orthogonal to our work.
Binary analysis application developers can benefit from our work and seek additional parallelization opportunities if necessary.

\textbf{Stripped binaries:} Stripped binaries do not have the static symbol table (\code{.symtab}) any more,
but still have the dynamic symbol table (\code{.dynsym}) and the exception unwinding frame information (\code{.eh\_frame}).
In addition, our algorithms can be augmented with orthogonal research for identifying stripped function entry points~\cite{Shin2015FEP, Bao2014BYTEWEIGHT,Rosenblum2008FEP}.

\textbf{Source code CFG construction:} The challenges of binary code CFG construction are largely distinct from those of source code CFG construction.
First, binary code functions can share code, which is the main reason that we must derive operation properties to guide our design invariants to support analysis of multiple functions in a binary in parallel. In contrast, source code functions cannot overlap unless functions are nested. In this case, CFG construction for source code does not require rigorous synchronization. For example, a source code basic block parsed by one thread is not going to be split by another thread.
Second, jump tables in binary code are often used to implement switch statements in the source code. Jump tables are encoded as indirect control flow in the binary code, whose targets must be identified through data flow analysis. However, in source code, the body of a switch statement is naturally grouped together, and it is straightforward to identify every case clause for the switch statement.
Third, the body of a source code function is contiguous. However, basic blocks of binary code functions can be outlined to improve instruction cache performance. As a result, binary analysis needs to address non-contiguous functions.
Fourth, tail calls in binary code are just normal function calls in source code.

\section{Conclusion}

With the increasing size of software and the need for analyzing large batches of binaries,
adding multithreaded parallelism speeds up binary analysis, but doing so
requires principled algorithm and data structure redesign
and careful attention to implementation.
Our work centers on a theoretical abstraction that expresses CFG construction as applications of individual CFG operations.
We derived operation dependencies, commutativity, and monotonic ordering properties,
which enable us to assess the strengths and weaknesses of existing serial CFG constructions,
and guided us towards a new design for our parallel CFG construction algorithm.
We evaluated our parallel binary analysis with a performance analysis tool \texttt{hpcstruct} and a software forensics tool \texttt{BinFeat},
achieving 25$\times$ speedup for parallel CFG construction, 14$\times$ for ingesting DWARF,
8$\times$ overall for \texttt{hpcstruct}, and 7$\times$ overall for \texttt{BinFeat} using 64 hardware threads.
Our results show that our parallel binary analysis can significantly speed up binary analysis applications, cutting the wait times for their users and developers.

%% Acknowledgments
%\begin{acks}                            %% acks environment is optional
                                        %% contents suppressed with 'anonymous'
  %% Commands \grantsponsor{<sponsorID>}{<name>}{<url>} and
  %% \grantnum[<url>]{<sponsorID>}{<number>} should be used to
  %% acknowledge financial support and will be used by metadata
  %% extraction tools.
%  This material is based upon work supported by the
%  \grantsponsor{GS100000001}{National Science
%    Foundation}{http://dx.doi.org/10.13039/100000001} under Grant
%  No.~\grantnum{GS100000001}{nnnnnnn} and Grant
%  No.~\grantnum{GS100000001}{mmmmmmm}.  Any opinions, findings, and
%  conclusions or recommendations expressed in this material are those
%  of the author and do not necessarily reflect the views of the
%  National Science Foundation.
%\end{acks}

%% Bibliography
\bibliography{pp}

\end{document}

%% file: config.tex
\usepackage{tikz}
\usepackage{amsmath}
\usepackage{amsfonts}
\usepackage{xcolor, colortbl}
\usepackage{collcell}
\usetikzlibrary{positioning}
\usetikzlibrary{arrows}
\usetikzlibrary{plotmarks}
\usepackage{pgfplots}
\usepackage{pgfplotstable}
\pgfplotsset{compat=1.14}
\usepackage{tabularx}
\usepackage{diagbox}
\usepackage{multirow}
\usepackage{caption}
\usepackage{enumitem}
\usepackage{bigdelim}
\usepackage{calc}
\usepackage{siunitx}
\usepackage{fancyhdr}
\usepackage{hyperref}
\usepackage[noabbrev,capitalize]{cleveref}
\usepackage{diagbox}
\usepackage{graphicx}
\usepackage{bm}
\usepackage{pifont}
\usepackage{subcaption}
\usepackage{adjustbox}

\crefrangelabelformat{Figure}{#3#1#4--#5\crefstripprefix{#1}{#2}#6}

\newcommand{\code}[1]{{\small\texttt{#1}}}

\newlength\figureheight
\newlength\figurewidth
\setlist{topsep=0pt,nolistsep}